\documentclass[10pt,notitlepage,pra,twocolumn]{revtex4-1}
\usepackage{amsfonts}
\usepackage{amsmath}
\usepackage{amssymb}
\usepackage{graphicx}
\usepackage{float}
\usepackage[toc,page,header]{appendix}
\usepackage{color}
\usepackage{multirow}

\begin{document}

\title{Switchable dynamic Rydberg-dressed excitation via a cascaded double electromagnetically induced transparency}
\author{Yichun Gao, Yinghui Ren, Dongmin Yu, Jing Qian$^\dagger$}
\affiliation{State Key Laboratory of Precision Spectroscopy, Department of Physics, School of Physics and Material Science, East China
Normal University, Shanghai 200062, China}

\begin{abstract}
Dynamic control of atomic dressing to the highly-excited Rydberg state in multi-level systems has special appeals owing to the development of flexible and precise measurement. In this study we develop an experimentally-accessible proposal to robustly control the dressing probability via a three-step cascaded excitation with double electromagnetically induced transparency (EIT) technique. The system can function as an optical switch where the third addressing laser serving as the control knob can switchably engineer the dressing probability with time. Differing from a conventional two-photon EIT, this novel scheme facilitates the maximal dressing probability determined by a relative strength between two coupling fields, entirely relaxing the absolute values for strong lasers. The collective feature caused by the interactions of a few atoms is also studied leading to an enhanced dressing probability as well as a reduced response time. Our work offers the oppotunity to a coherent dynamic control of Rydberg excitation and to realize sizable Rydberg-Rydberg interactions in weakly-driven quantum systems.
\end{abstract}

\maketitle
\preprint{}

\section{Introduction}
Rydberg dressed atoms working in a regime with only a small fraction of excited atoms, promises the sizable long-range interaction to megahertz as well as the enhanced lifetime up to a few seconds \cite{Johnson10,Saffman16}. They have great prospectives for the studies of strongly correlated many-body physics \cite{Honer10,Pupillo10,Henkel10,Balewski14}, the laser-driven quantum magnetism \cite{Bijnen15,Glaetzle15} and the micro-devices for quantum information processing \cite{Keating13,Keating15,Jau16}. The route to Rydberg dressing can be accessible via Rydberg electromagnetically induced transparency (Rydberg-EIT) \cite{Mohapatra07,Pritchard10,Petrosyan11} which stimulates resonant optical driving by the establishment of a dark state weakly dressed to a Rydberg level \cite{Gaul16}, opening up new research avenues for nonlinear quantum optics \cite{Firstenberg16}.

The optical control of resonant dipole-dipole interactions has been realized in a four-level protocol with two upper Rydberg levels coupled by a microwave field, inducing exchange interaction between them \cite{Vogt18}. Such a multi-step scheme to highly-excited Rydberg states is sufficiently appealing because it enables a Rydberg state to be easily reached by nearby levels with inexpensive lasers \cite{Leseleuc17,Carr12}. Besides, it experimentally benefits from a kick-free and Doppler-free excitation that can produce a narrow resonance linewidth. So far, a few pioneer works studying EIT effect in a four-level atomic system have originated from Agarwal \cite{Agarwal96} and Harris \cite{Harris98}, describing the absorption of two photons controlled by quantum interference between the two excitation paths. Later, a number of theoretical works studying the nonlinear optical feature are represented in similar systems \cite{Mulchan00,Dutta07,Che14}. Recently a more complex four-step excitation scheme is achieved in which the uppermost Rydberg state has a large dipole moment \cite{Kondo15}.

However, most studies have been limited to the steady-state property of the probe field, revealing stable transmission signals in {\it{e.g.}} hollow-core fibers \cite{Veit16}; a coherent dynamic control for the atomic dressing to a Rydberg level is still rarely reported, mainly caused by the complexity of multiple lasers in multi-level systems.
Recently {\it Zhang etc.} studied the transient probe absorption response of a Rydberg-EIT by a sudden switch-on of the coupling field, bringing one-step closer to this target \cite{Zhang18}; while a complete study of dynamic control and response remains elusive.

In the present work, we propose to study practical dynamic dressing process to a target Rydberg level with a three-step cascaded excitation. 
A usual two-step EIT can couple a ground state to a long-lived $nS$ or $nD$-type state by the transition rules\cite{Ate11,Xu16,Sheng17,Jiao17}. Here the uppermost state is $nP$-type, one has to use at least a three-step cascaded excitation where the third addressing field modified to be a time-dependent square-wave pulse serving as a control knob, results in a split for the probe field transmission window, naming as double Rydberg-EIT \cite{Rawat18}. In spite of the complexity of system, we observe that the maximal dressing probability is only sensitive to the relative strength between two strong coupling lasers while their absolute values are nearly irrelevent. 
A generalized constraint for realizing an optimal dressing probability is proposed, which is  $\Omega_{20}/\Omega_1\approx\sqrt{\gamma/\Gamma}$ and $\Omega_1/\Omega_p\gg1$ with the laser amplitudes $\Omega_{20}$, $\Omega_1$, $\Omega_p$.

In addition, we study the transient response in the system and describe the role of Rydberg blockade in the reduction of response time, accompanied by an enhanced dressing probability due to the collective feature on dark-state resonance by interacting atoms \cite{Chai17}. The effect is improved with the number of atoms. An experimental feasibility is discussed by switchably turning on or off the addressing field with fullly flexible parameters, demonstrating the realization of a robust atomic dressing excitation by saving the laser powers on demand in the implementation. Our results provides a new treatment to simulate the process of multi-level Rydberg dressing with optimal dressing probability and response time, facilitating its extensive uses in developing all-optical switches and transistor devices \cite{Baur14,Gorniaczyk14,Li15,Qian17}.

\section{Theoretical model and steady state}

\begin{figure}
\includegraphics[width=3.4in,height=3.3in]{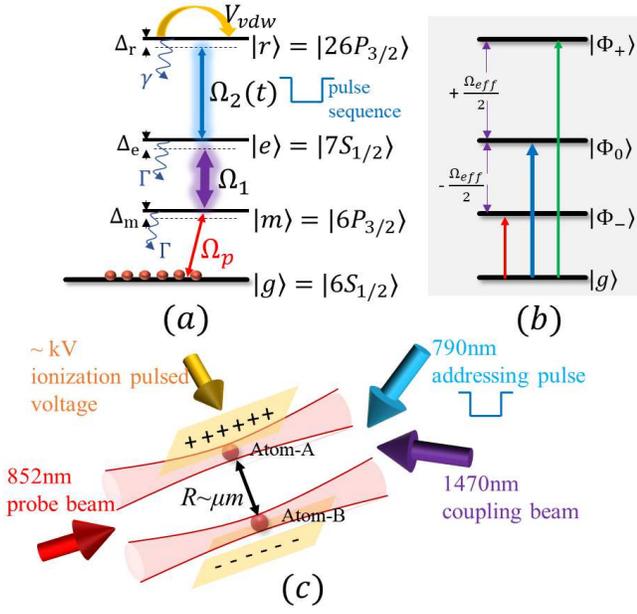}
\caption{Schematics of a controlled three-step Rydberg dressing excitation. (a) Energy levels and the atom-field interactions. Ground-state atoms are dressed to a Rydberg state $|r\rangle$ via a cascaded three-step process $|g\rangle\to|m\rangle\to|e\rangle\to|r\rangle$. Real energy levels are shown based on Cs atoms. (b) Dressed bright eigenstates $|\Phi_0\rangle$, $|\Phi_{\pm}\rangle$ composing a small fraction of $|r\rangle$ under the condition of $\Omega_1,\Omega_{20}\gg\Omega_p$ and their couplings with the dark-state $|g\rangle$ are demonstrated. (c) The experimental setup (proposal). Two Cs atoms are trapped in tightly focused dipole traps with the interstate spacing $R\sim\mu m$, detected by using a pulse field ionization with a high voltage $\sim kV$.}
\label{mod}
\end{figure}

As displayed in Fig.\ref{mod}(a) our model consists of a cascaded three-step excitation to the Rydberg state $|r\rangle$ following $|g\rangle\to|m\rangle\to|e\rangle\to|r\rangle$, performed by one weak probe laser $\Omega_p$, one strong coupling laser $\Omega_1$ and one addressing coupling laser $\Omega_2(t)$. Specially, the addressing laser $\Omega_2(t)$ with amplitude $\Omega_{20}$ is a time-dependent pulse sequence acting as a control knob to the process. The system can be described by a double-EIT scenario where the third laser with detuning $\Delta_r=0$ results in a split transparent resonance for the probe absorption by dark-state resonance with the three bright dressed states $|\Phi_{0,\pm}\rangle$ [Fig.\ref{mod}(b)]. In the absence of $\Omega_2$ it works as a conventional EIT between $|g\rangle$ and $|e\rangle$, suppressing the dressing probability on state $|r\rangle$. An experimental setup is presented in Fig.\ref{mod}(c) where the three-step excitation is played by the lasers with wavelengths 852nm, 1470nm, and 790nm, respectively. The measurement can be performed by using an ultra-short high-voltage electric pulse for ionization of Rydberg state atoms, accompanied by a time of flight spectra detecting ions or electrons \cite{Zhang09}. 

In the interaction frame the Hamiltonian $\mathcal{H}_N$ for one atom can be described by ($\hbar=1$)
\begin{equation}
\hat{\mathcal{H}}_{N}=\hat{\mathcal{H}}_{N,0}+\dfrac{1}{2}(\Omega_p\hat{\sigma}_{N,gm}+\Omega_1\hat{\sigma}_{N,me}+\Omega_{20}\hat{\sigma}_{N,er}+H.c.)
\label{Ham2}
\end{equation}
where $\hat{\mathcal{H}}_{N,0}=-[\Delta_{m}\hat{\sigma}_{N,mm}+(\Delta_{m}+\Delta_{e})\hat{\sigma}_{N,ee}+(\Delta_{m}+\Delta_{e}+\Delta_{r})\hat{\sigma}_{N,rr}]$ describes the atomic self-energy under resonance: $\Delta_e=\Delta_r=0$. Here $\hat{\sigma}_{N,\alpha\beta}=|\alpha\rangle_N\langle \beta|$ is the projection operator. The subscript {\it N}=1 means the case of one atom. In experiment the detuning $\Delta_m$ can be easily tuned by the probe laser frequency for observing the probe transmission. $\Gamma$ and $\gamma$ are the spontaneous decay rates of middle excited $|m(e)\rangle$ and Rydberg levels $|r\rangle$, typically $\gamma/\Gamma=0.01$.

The steady state solutions can be analytically solved.
For $\Omega_p\ll\Omega_1,\Omega_{20}$ (we define $\Omega_{eff}=\sqrt{\Omega_1^2+\Omega_{20}^2}$), the subspace with $|m\rangle$, $|e\rangle$, $|r\rangle$ is composed of three dressed bright states $|\Phi_0\rangle$ and $|\Phi_{\pm}\rangle$, given by
\begin{eqnarray}
|\Phi_0\rangle&=&\frac{1}{\Omega_{eff}}(-\Omega_{20}|m\rangle+\Omega_1|r\rangle) \label{p0}\\
|\Phi_{\pm}\rangle&=&\frac{1}{\sqrt{2}\Omega_{eff}}(\Omega_1|m\rangle\pm\Omega_{eff}|e\rangle+\Omega_{20}|r\rangle) \label{p+}
\end{eqnarray}
with eigenvalues $E_{\Phi_0}=-\Delta_m$, $E_{\Phi_{\pm}}=\frac{1}{2}(-2\Delta_m\pm\Omega_{eff})$, which are coupled to the ground dark state $|D\rangle=|g\rangle$ with eigenvalue $E_{0}=0$. Note that $|\Phi_{\pm}\rangle$ is unsuited for atomic dressing due to its non-negligible occupancy on state $|e\rangle$, {\it i.e.} $P_{1,e}\equiv \frac{1}{2}$; however if $P_{1,m}=(\Omega_{20}/\Omega_{eff})^2\ll1$, $|\Phi_{0}\rangle$ is a good candidate for dressing based on its resonance with $|D\rangle$ at $\Delta_m=0$ as well as its complete immunity to the middle excited state $|e\rangle$.

The steady probe absorption and the Rydberg dressing probability can be directly solved from the master equation: 
\begin{equation}
\frac{d\hat{\rho}}{dt}=i[\hat{\rho},\hat{\mathcal{H}}_1]+\hat{\mathcal{L}}_1(\hat{\rho})
\label{maseq}
\end{equation}
where the Lindblad operator taking into account dissipative processes can be demonstrated as
\begin{eqnarray}
\mathcal{\hat{L}}_1(\hat{\rho})&=&\Gamma \sum_{\alpha\beta}[\hat{\sigma}_{1,\alpha\beta}\hat{\rho}\hat{\sigma}_{1,\beta\alpha}-\frac{1}{2}(\hat{\sigma}_{1,\beta\beta}\hat{\rho}+\hat{\rho}\hat{\sigma}_{1,\beta\beta})] \nonumber \\ 
&+&\gamma[\hat{\sigma}_{1,er}\hat{\rho}\hat{\sigma}_{1,re}-\frac{1}{2}(\hat{\sigma}_{1,rr}\hat{\rho}+\hat{\rho}\hat{\sigma}_{1,rr})]
\end{eqnarray}
with $\alpha\beta\in\{gm,me\}$. In the weak probe limit where all population is assumed to be in the ground state by letting $ \rho_{gg}^{s}=1, \rho_{mm}^{s}=\rho_{ee}^{s}=\rho_{rr}^{s}=0$ (the superscript {\it s} means the steady state), the non-diagonal steady element $\rho_{gm}^s$ under $\dot{\rho}_{ij}=0$ demonstrating the coherence of transition, can be expressed as
\begin{flalign}
\rho_{gm}^{s}=i\Omega_{p}\dfrac{\Omega_{20}^{2}+X_{1}X_{2}}{X_{1}\Omega_{20}^{2}+X_{2}\Omega_{1}^{2}+X_{1}^{2}X_{2}} \label{gm}
\end{flalign}
where $X_{1}=\Gamma+2i\Delta_{m}$, $X_{2}=\gamma+2i\Delta_{m}$ and $\Gamma_{m}=\Gamma_{e}=\Gamma, \Gamma_{r}=\gamma$. The probe absorption and dispersion can be solved from the imaginary and real part of $\rho_{gm}^s$, and the steady dressing probability $\rho_{rr}^s$ labeled by $P_{1,r}$ to the first-order approximation takes a complex form as represented in the Appendix A.

 \begin{figure}
\includegraphics[width=3.2in,height=3.5in]{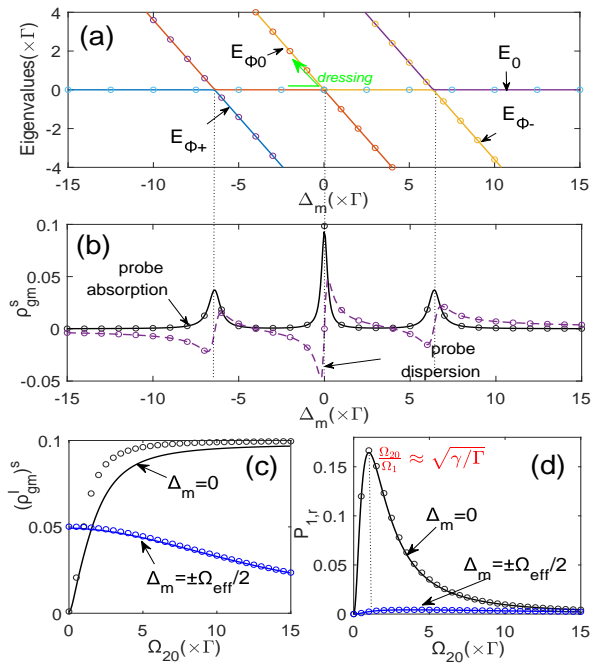}
\caption{(a) The eigenvalues and (b) the enhanced steady probe absorption $(\rho_{gm}^I)^s$ and dispersion $(\rho_{gm}^R)^s$ versus $\Delta_m$. Three absorption peaks emerge at the avoided crossings $\Delta_m=0,\pm\Omega_{eff}/2$. Parameters used are $\Omega_1/\Gamma=10$, $\Omega_{20}/\Gamma=8$, $\Omega_p/\Gamma=0.1$, $\gamma/\Gamma=0.01$. The peak probe absorption strength $(\rho_{gm}^I)^s$ in (c) and the steady dressing probability $P_{1,r}$ in (d) at $\Delta_m=0$ (black) and $\pm\Omega_{eff}/2$ (blue) by adjusting $\Omega_{20}$. Numerical and analytical results are represented by solid curves and circles, respectively. }
\label{eigsp}
\end{figure}

For an intuitive understanding, relevant eigenvalues with respect to $\Delta_m$ are comparably plotted in Fig.\ref{eigsp}(a). It is clear that there exists an zero-energy dark state $E_0$, avoidedly crossed by $E_{\Phi_0}$ and $E_{\Phi_{\pm}}$ at $\Delta_m=0, \pm\Omega_{eff}/2$, arising enhanced probe absorption there, see Fig.\ref{eigsp}(b). The peak probe absorption labeled by $(\rho_{gm}^I)^s$ as well as the Rydberg dressing probability $P_{1,r}$ at $\Delta_m=0, \pm\Omega_{eff}/2$ versus $\Omega_{20}/\Gamma$ are also represented in Fig.\ref{eigsp}(c-d). In a conventional EIT scenario with $\Omega_{20}=0$ the probe laser experiences no absorption in the center($\Delta_m=0$), arising zero excitation probability there. By increasing $\Omega_{20}$ the central resonant absorption induced by the constructive quantum interference between $|g\rangle\to|\Phi_+\rangle$ and $|g\rangle\to|\Phi_-\rangle$ obtains a substantial growth, accompanied by a slight reduction for the off-resonant absorptions at $\Delta_m=\pm\Omega_{eff}/2$. 
All analytical expressions labeled by circles in Fig. \ref{eigsp} perfectly agree with our numerical simulations.

For improving the dressing probability it requires a direct connection between $|g\rangle$ and $|\Phi_{0}\rangle$ at the avoided crossing, see Fig.\ref{eigsp}a(green arrow). From Eq.(\ref{p0}), the occupancy in $|m\rangle$ must be minimized by $P_{1,m}=(\Omega_{20}/\Omega_1)^2\ll 1$, agreeing with the finding in Fig.\ref{eigsp}d that the optimal dressing probability $P_{1,r}^{opt}$ occurs at $\Omega_{20}/\Omega_1\approx0.1$(quite small) because of the assumption of $\gamma/\Gamma=0.01$. In fact, via a differential calculation to Eq.(A3) we obtain a generalized condition 
\begin{equation}
\Omega_{20}/\Omega_1\approx \sqrt{\gamma/\Gamma}
\end{equation}
by which an optimal value $P_{1,r}^{opt}\approx0.16$ is achievable with the constraint that $\Omega_p\ll\Omega_1$. In other words, an optimal dressing probability in our model is only determined by the relative strength between two strong coupling lasers $\Omega_1$, $\Omega_{20}$, rather than their absolute values, providing a more flexible selection of lasers in the implementation.
Additionally, note that if $\Omega_{20}$ is even larger, for $\Delta_m=0$ $P_{1,r}$ has a quick decrease owing to the effect of $|m\rangle$ while at off-resonance $P_{1,r}$ is kept to be zero due to its poor dressing probability with $|\Phi_{\pm}\rangle$. To this end we will  focus on the case of $\Delta_m=0$.

\section{Dynamic atomic dressing to a Rydberg state}
 
In experiment, a study of transient behavior for dynamic dressing to a Rydberg state is significantly important, requiring a sensitive measurement. Here we theoretically investigate the dynamic dressing excitation of system with a time-dependent pulse $\Omega_2(t)$, simluating switch-off and switch-on cases with respect to $\Omega_2(t)=0$ and $\Omega_2(t)=\Omega_{20}$. Intuitively, when $\Omega_2(t)$ is turned off, it completely closes the excitation and if $\Omega_2(t)$ is turned on, the dressing dynamics experiencing a fast response speed ($\sim\mu s$), arrives at a new higher steady state, see the inset of Fig.\ref{sig}b. Numerical results are obtained by directly solving the optical Bloch equations (A1) in the Appendix A where the dephasing effects acting on coherence due to transit time broadening and laser intensity variations are ignored. 

The performance of such an optical switch can be quantified by two important parameters $P_{1,r}^{on}$ and $\tau_{1,r(d)}$,

(i) The excitation probability $P_{1,r}^{on}$ with the superscript {\it on} for the switch-on case. Note that $P_{1,r}^{off}\equiv0$ due to $\Omega_2(t)=0$ in the switch-off case. Greatly differing from a two-photon EIT, the middle laser $\Omega_1$ brings in an auxiliary adjustment, allowing $P_{1,r}^{on}\to P_{1,r}^{opt}$ with flexible parameters.

(ii) The transient response speed $\tau_{1,r(d)}$ represents the required time for attaining a new steady state after a sudden switch of $\Omega_2(t)$. $\tau_{1,r}$ and $\tau_{1,d}$ respectively stand for the rise time and the fall time and the subscript $1$ means one atom.

\begin{figure}
\includegraphics[width=3.4in,height=2.8in]{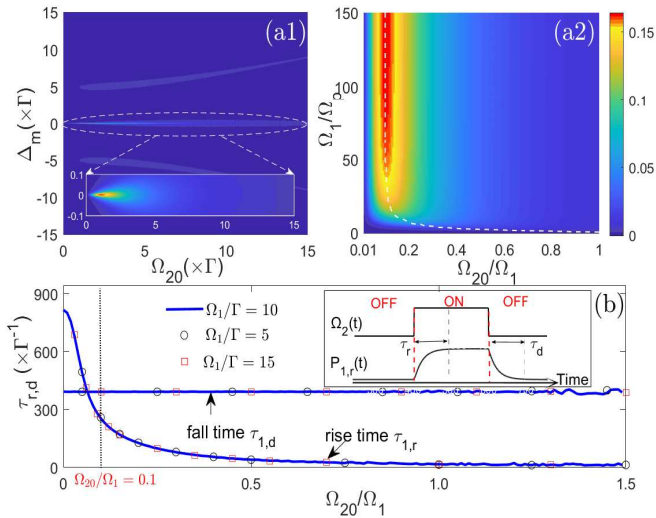}
\caption{(color online) In switch-on case, the excitation probability $P_{1,r}^{on}$ versus the variations of amplitudes $\Omega_{20}$ and $\Delta_m$ in (a1), as well as of ratios $\Omega_{20}/\Omega_1$ and $\Omega_{1}/\Omega_{p}$ in (a2). Inset of (a1) is an enlarged picture for $\Delta_m/\Gamma\in[-0.1,0.1]$. The local maximal values with respect to given $\Omega_{1}/\Omega_{p}$ values are denoted by a white dashed curve. (b) The transient response time $\tau_{1,r(d)}$ (rise time and fall time) versus the change of ratio $\Omega_{20}/\Omega_1$ for $\Omega_1/\Gamma=10$ (blue solid), 5 (black circles), 15 (red squares). It keeps $\Omega_p/\Gamma=0.1$. Inset of (b) shows a complete one-period dynamics of Rydberg dressing probability $P_{1,r}(t)$ under a controlled pulse sequence $\Omega_2(t)$. Here $\gamma/\Gamma=0.01$. }
\label{sig}
\end{figure}

As represented in Fig.\ref{eigsp}d the dressing probability $P_{1,r}^{on}$ has a significant growth on resonance $\Delta_m=0$, agreeing with the results of Fig.\ref{sig}(a1) where we plot $P_{1,r}^{on}$ versus $\Delta_m$, $\Omega_{20}$, and an obvious enhancement on resonance is verified. So we only consider $\Delta_m=0$ later. An enlarged view in the inset of (a1) reveals a similar behavior as Fig.\ref{eigsp}d(black solid) that $P_{1,r}^{on}$ has a continuously decreasing after its peak value at $\Omega_{20}/\Omega_1\approx\sqrt{\gamma/\Gamma}$, strongly proving the importance of ratio $\Omega_{20}/\Omega_1$ for realizing an optimal value $P_{1,r}^{opt}$. 
To improve the dressing probability letting $P_{1,r}^{on}\to P_{1,r}^{opt}$, in Fig.\ref{sig}(a2) we globally study the dependence of $P_{1,r}^{on}$ on relative optical Rabi frequencies $\Omega_1/\Omega_p$, $\Omega_{20}/\Omega_1$. It is remarkably noting that the dressing probability robustly sustains it optimal value by satisfying $\Omega_{20}/\Omega_1\approx\sqrt{\gamma/\Gamma}=0.1$ once $\Omega_1/\Omega_p\gg1$ is met. Here $\Omega_1/\Omega_p\geq50$ is the least requirement. In other words the absolute values of optical Rabi frequencies are fully independent in our scheme, providing us a more flexible way to realize an improving Rydberg dressing probability with much reduced laser powers.

For gaining a deep and complete insight, we further study the transient response speed by dynamically switching on or off the addressing laser $\Omega_2(t)$, and record the required responce times for a new steady state after the switches. Figure \ref{sig}(b) represents the fall time $\tau_{1,d}$ and the rise time $\tau_{1,r}$ as a function of $\Omega_{20}/\Omega_1$ for different $\Omega_1$. Note that here the condition $\Omega_1/\Omega_p\gg1$ is always preserved. Surprisingly, the fall time $\tau_{1,d}$ is observed to be kept a constant irrespective of the laser drivings because when $\Omega_2(t)$ is switched off the system always tends to a same steady state with $P_{1,r}^{off}= 0$, arising $\tau_{1,d}$ decided by the decay rates only. If one reduces the value $\gamma$ by exciting to a higher-$n$ Rydberg state, it needs more time to be stable, see Table \rm{I} for a detailed discussion. In contrast, the rise time $\tau_{1,r}$ exhibits a rapid decrease with the increase of $\Omega_{20}/\Omega_1$, essentially caused by a stronger excitation between $|e\rangle$ and $|r\rangle$ with a larger $\Omega_{20}$ or a smaller $\Omega_1$, reducing the time for attaining stable. Also it needs to stress again that only the relative strengths of optical Rabi frequencies not their absolute strengths play roles, because different $\Omega_1$ values give rise to exactly same response times in our proposal.

In Appendix B we give more informantion for comparing the response speeds between the probe absoption and dressing dynamics. To our knowledge, a sudden switch of the addressing field $\Omega_2(t)$ can cause a transient optical response for the probe field absorption, subsequently arising a tendency towards a new steady state within a small time. The response time for the probe absorption is predicted to be faster than for the yield of steady dressing probability. In our calculations the response speed for dressing process is $(1\sim10)\mu$s, and for the probe absoption is only $(0.1\sim 1.0)\mu$s

To briefly summarize, based on a four-level cascaded system, we find that the optimal dressing probability $P_{1,r}^{opt}[\approx0.16]$ is only determined by a relative strength between two strong coupling fields $\Omega_1$ and $\Omega_{20}$, rather than their absolute values when $\Omega_1/\Omega_p\gg1$ is met, offering a great challenge to save the laser powers in an experimental implementation. However we also note that at $\Omega_{20}/\Omega_1\approx\sqrt{\gamma/\Gamma}$ the rise time $\tau_{r}$ is relative longer $\sim 300\Gamma^{-1}$, see black dotted line in Fig.\ref{sig}b, which can be improved by considering a few interacting atoms, promoting us to study the effect of collective feature in the next section.

\section{sizable Rydberg-Rydberg interactions}

In this section we investigate how the van der Waals(vdWs) interaction affects the transient dressing dynamics in a few atomic system.
To our knowledge, a two-atom system is the simplest model to study the collective feature that has been widely achieved in experments \cite{Urban09,Gaetan09,Wilk10}. For that reason we adopt two atoms with tunable vdWs interactions denoted as $V_{vdW}$, performed by varying the interatomic spacing $R$. For the interaction strength much larger than Rabi frequencies, the collective Rabi oscillation will be enhanced by a factor of $\sqrt{2}$, leading to an enhancement for the dressing probability \cite{Heidemann07}; however it may fail due to the breakup of the blockaded condition \cite{Beguin13} or the resonance by anisotropic interactions \cite{Qian16}.

Presently, we employ an approach of the two-atom (A and B) master equation where the Hilbert space is expanded by full $4^2$ basis vectors, giving rise to a $4^2\times4^2$ density matrix $\rho$, governed by
\begin{equation}
\frac{d\hat{\rho}}{dt}=i[\hat{\rho},\mathcal{\hat{H}}_{2}]+\mathcal{\hat{L}}_2(\hat{\rho})
\label{mas}
\end{equation}
with $\mathcal{\hat{H}}_{2}=\mathcal{\hat{H}}_{2}^A+\mathcal{\hat{H}}_{2}^B+\mathcal{\hat{H}}_{2}^{af}$ and $\mathcal{\hat{H}}_{2}^{af}=V_{vdW}\hat{\sigma}_{2,rr}^A\hat{\sigma}_{2,rr}^B$ describes the vdWs interaction between two Rydberg atoms (the superscripts ``A" and ``B" refer to different atoms), and $\mathcal{\hat{L}}_2(\hat{\rho})= \Gamma\sum_{\alpha\beta}\sum_{i=A,B}[\hat{\sigma}_{2,\alpha\beta}^i\hat{\rho}\hat{\sigma}_{2,\beta\alpha}^i-\frac{1}{2}(\hat{\sigma}_{2,\beta\alpha}^i\hat{\sigma}_{2,\alpha\beta}^i\hat{\rho}+\hat{\rho}\hat{\sigma}_{2,\beta\alpha}^i\hat{\sigma}_{2,\alpha\beta}^i)]+\gamma\sum_{i=A,B}[\hat{\sigma}_{2,er}^i\hat{\rho}\hat{\sigma}_{2,re}^i-\frac{1}{2}(\hat{\sigma}_{2,re}^i\hat{\sigma}_{2,er}^i\hat{\rho}+\hat{\rho}\hat{\sigma}_{2,re}^i\hat{\sigma}_{2,er}^i)]$ is the Lindblad operator 
with $\alpha\beta\in\{gm,me\}$, presenting the effect of spontaneous decays. 
For comparison we also adopt two important parameters, denoted by $P_{2,r}^{on}$ describing the singly-excited-state dressing probability in the presence of $\Omega_2$ as well as $\tau_{2,r(d)}$ for the response speed of $P_{2,r}(t)$. Intuitively, $P_{2,r}^{on}$ is predicted to be enhanced by $\sqrt{2}$ with respect to $P_{1,r}^{on}$ by the effect of collective excitation. In fact, this factor can exceed $\sqrt{2}$ in multi-level systems with comparable Rydberg interactions \cite{Ma19}. For comparison, numerical results based on three interacting atoms are simultaneously shown.

\begin{figure}
\includegraphics[width=3.4in,height=2.8in]{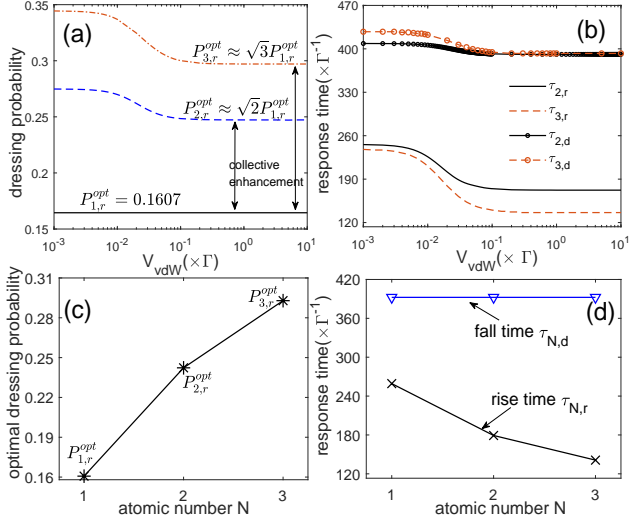}
\caption{(a) The singly-excited-state dressing probabilities $P_{3,r}^{on}$ (red dash-dotted), $P_{2,r}^{on}$ (blue-dashed) and $P_{1,r}^{opt}$ (black-solid) for different atoms $N=3,2,1$ versus $V_{vdW}$. (b) The rise time $\tau_{N,r}$(solid and dashed curves) and the fall time $\tau_{N,d}$ (solid and dashed curves with circles) of atom $N$ versus $V_{vdW}$. (c-d) The optimal dressing probability $P_{N,r}^{opt}$ and the required response time $\tau_{N,r(d)}$ versus the number $N$ under strong blockade.}
\label{tat}
\end{figure}

Numerical solving Eq.(\ref{mas}) gives rise to an observable measurement for the singly-excited-state dressing probability $P_{2,r}(t)$ as well as the response behavior $\tau_{2,r(d)}$. Here we pay attention to the dressing probability $P_{2,r}^{on}$ under the switch-on case, and a complete time-dependent output will be studied in section \rm{V}. In section \rm{III} we have found an optimal dressing probability of single atom, which is $P_{1,r}^{opt}\approx0.1607$ by satisfying a generalized condition of $\Omega_{20}/\Omega_1=\sqrt{\gamma/\Gamma}$ and $\Omega_1/\Omega_p\gg1$. 
Turning to the picture of two or more interacting atoms, due to the collective feature induced by strong interatomic interactions, $P_{2,r}^{opt}$ and $P_{3,r}^{opt}$ reveal an explicit improvement whose enhancement factor approximately equals to $\sqrt{N}$ ($N$ is the number of atoms) in the strong blockade regime as represented in Fig.\ref{tat}a. In the weak blockade regime $P_{N,r}^{opt}$ is estimated to be bigger due to definition $P_{2,r}^{opt}=P_{1,r}^{A}+P_{1,r}^{B}-2P_{2,rr}^{AB}$ that a strong blockaded interaction would reduce the probability of one of two atoms' excitation, making the value $P_{2,r}^{opt}$ smaller. Fig.\ref{tat}c gives a full view of optimal singly-excited-state dressing probability $P_{N,r}^{opt}$ with the atomic number $N$, supporting a robust enhancement.

Correspondingly, in Figure \ref{tat}b we also explore the response speeds $\tau_{2,r(d)}$, $\tau_{3,r(d)}$ by the effect of vdWs interaction. The collective feature will induce a faster-responding dynamic property. As increasing $V_{vdW}$ both the rise and fall times reveal an observable decrease. The physical description for a faster response time is that a stronger interaction strength makes the Rydberg state out of resonance, allowing the relaxation process mainly affected by a big decay from the middle excited states. However, $\tau_{2,r(d)}$ tends to be unchanged when the interaction strength is sufficiently strong that the excitation of two atoms is completely blockaded. Fig.\ref{tat}d shows the response time under full blockade versus the increase of atomic number, and it is found that the fall time $\tau_{N,d}$ is kept to be a constant as similar as predicted by Fig.\ref{sig}b, that is only determined by the Rydberg decay $\gamma$. More interestingly the rise time $\tau_{N,r}$ shows an obvious decrease with $N$, providing more prospectives with faster response time and more efficient dressing probability in a $N$-atom ensemble.

\section{Optimal experimental implementation}

A practical dynamic dressed-atom generation via coherent control pulses can be considered using Cs atoms with relevant states $|g\rangle=|6S_{1/2}\rangle$, $|m\rangle=|6P_{3/2}$, $|e\rangle=|7S_{1/2}\rangle$, $|r\rangle=|26P_{3/2}\rangle$, carried out by an experimental setup like Fig.\ref{mod}c. In the calculations some parameters are estimated to be constant {\it i.e.} $\Gamma/2\pi=6.0$MHz, $\Omega_p/2\pi=0.6$MHz \cite{Beterov09}. For two- or more-atom case, we assume a strong blockaded interaction by considering $C_6/2\pi=140$GHz$\mu$m$^6$ and the interatomic spacing $R=3.64\mu$m, giving rise to $V_{vdW}/2\pi=60$MHz \cite{Zoubi15}.

For a comparable representation, in Figure \ref{realp} we study the real dynamic output of atomic dressing probability $P_{N,r}(t)$ to the singly-excited Rydberg state by considering different Rydberg decay rates: $\gamma/2\pi=60$kHz (left panels) and $\gamma/2\pi=30$kHz (right panels). For $\gamma/2\pi=60$kHz it is verified that the optimal dressing probability $P_{N,r}^{opt}$ occurs at $\Omega_{20}/\Omega_1=\sqrt{\gamma/\Gamma}=0.1$ with the constraint that $\Omega_1/\Omega_p\geq50$ (i.e. $\Omega_1\geq30$MHz), by which exactly same probabilities are shown see (a1-a2). Compared to (a1), the results in (a2) benefits from an even lower laser Rabi frequency under the parameter optimization. However once one of two constraints breaks, the dressing probability will have a big reduction, see (a3) and (a4), strongly supporting the importance of two generalized constraints.

For comparison, while lowering the decay rate to $\gamma/2\pi=30$kHz it is observed that the required laser powers for a same output value can be further reduced even by one order of magnitude [see (a4) and (b4)]. That means, exciting to a higher-$n$ Rydberg level can save more laser powers which is more favorable by experimentalists. However, we also note that for a small $\gamma$ the system suffers from a slightly longer fall time $\tau_{N,d}$, as demonstated in section \rm{III}.

Specific parameters in (a2) and (b2) are summaried in Table \ref{ta} presenting a quantitative comparison for the realization of optimal dressing probability and corresponding responce speed in the scheme. In plotting Figure \ref{realp}(a2) we have used the least but optimal constraint for dressing probability, that is $\Omega_{20}/\Omega_1=0.1$ and $\Omega_1/\Omega_p=50$, leading to the maximal dressing probability $P_{N,r}^{opt}=0.1607,0.2423,0.2929$ respectively for atoms $N=1,2,3$. Obviously, with the increase of $N$ the collective feature of strongly interacting atoms allows the rise time to be significantly reduced from 6.87$\mu$s to $3.74\mu$s while the fall time keeps unvaried $\sim 10.398\mu$s, comparably to the experimental data [$\sim\mu s$] for collective Rabi oscillation and excitation dynamics with Rydberg states \cite{Johnson08,Dudin12,Zeiher15}. By decreasing $\gamma$ to $2\pi\times30$kHz, for realizing same maximal dressing probabilities, the required coupling laser $\Omega_1$ can even reduced to $2\pi\times17.7$MHz.

Finally it is remarkable that, compared to the laser Rabi frequencies typically scaled by $\sim 100$MHz for a cascaded excitation in current Rydberg experiments \cite{Sibalic16}, we propose a new scheme that can save the laser intensities to be $(0.1\sim10)$MHz on average, accompanied by an optimation for the transient response behavior, robustly facilitating its future use in preparation of higher-$n$ Rydberg atoms with cascaded multi-level systems.

\begin{widetext}

\begin{figure}
\includegraphics[width=4.9in,height=3.2in]{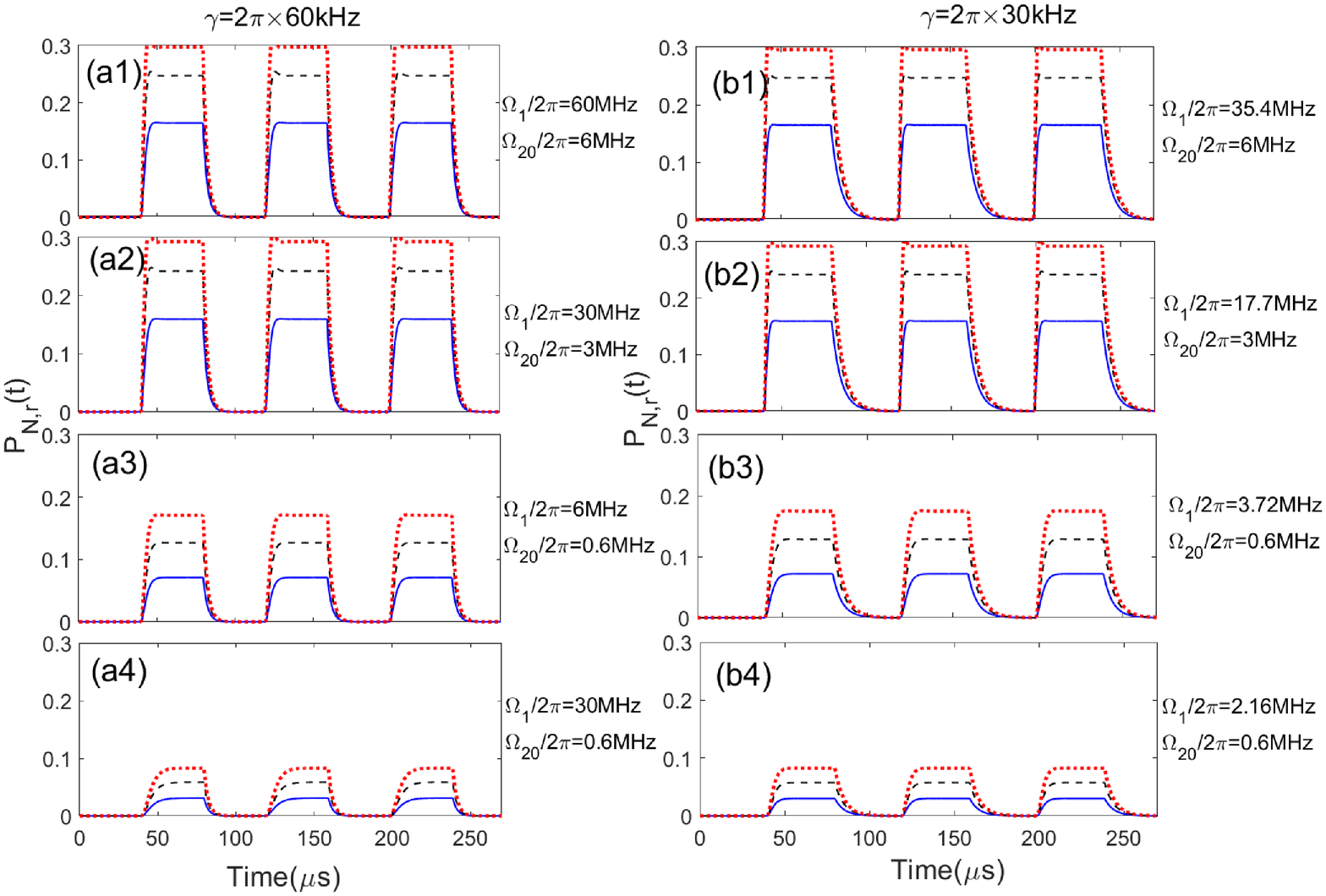}
\caption{(color online) Realistic performance for time-dependent singly-excited dressing probabilities $P_{N,r}$ of different atomic numbers $N=1$(blue solid), $2$(black dashed), $3$(red dotted). The required laser Rabi frequencies are listed on the right of panels and the corresponding Rydberg decay rate is shown on the top. Other parameters are $\Gamma/2\pi=6$MHz, $\Omega_p/2\pi=0.6$MHz, $\Delta_m=0$.
 }
\label{realp}
\end{figure}

\begin{table}[tbp]
\begin{tabular}{c|c|c|c|c|c|c}
\hline\hline
Rydberg decay & \multicolumn{2}{|c|}{key parameters} & atomic number & \multicolumn{2}{|c|}{response speed} & optimal dressing probability \\
\hline 
 $\gamma$($2\pi\times$kHz) & $\Omega_1$($2\pi\times$MHz) & $\Omega_{20}$($2\pi\times$MHz) & $N$ & $\tau_{N,r}(\mu s)$ & $\tau_{N,d}(\mu s)$ & $P_{N,r}^{opt}$  \\ 
 \hline 
 \multirow{3}*{60} & \multirow{3}*{30} & \multirow{3}*{3}
& 1 & 6.870$\mu s$ & 10.398$\mu s$ & 0.1607 \\ \cline{4-7}
 & & & 2 & 4.748$\mu s$ & 10.398$\mu s$ & 0.2423 \\ \cline{4-7}
 & & & 3 & 3.740$\mu s$ & 10.398$\mu s$ & 0.2929 \\
\hline
\multirow{3}*{30}
& \multirow{3}*{17.7}
& \multirow{3}*{3}
& 1 & 4.350$\mu s$ & 20.770$\mu s$ & 0.1607 \\ \cline{4-7}
 & & & 2 & 3.024$\mu s$ & 20.770$\mu s$ & 0.2418 \\ \cline{4-7}
 & & & 3 & 2.361$\mu s$ & 20.770$\mu s$ & 0.2923 \\
\hline\hline
\end{tabular}%
\caption{Numerical results with respect to Fig.\ref{realp}(a2) and (b2) are quantitatively given for comparing the optimal dressing probability $P_{N,r}^{opt}$ and its corresponding response speeds $\tau_{N,r(d)}$. By reducing the decay rate $\gamma$ one can realize same values of dressing probability with a lower laser power of $\Omega_1$, however arising a longer fall time $\tau_{N,d}$ at the same time. The rise time $\tau_{N,r}$ is observed to be shorter by decreasing $\gamma$ due to the growth of $\Omega_{20}/\Omega_1$.}%
\label{ta}
\end{table}

\end{widetext}

\section{Conclusions}

We propose a four-level cascaded scheme for switchable collective dressing excitation using a double Rydberg-EIT where the control operation is played by the third addressing laser with time-dependent property that directly connects to the Rydberg level. Special attention is paied on the optimization of dressing probability that a generalized condition to an optimal value is attainable which exactly depends on the relative strength of two strong coupling fields. Besides we also explore the transient responses of system by a sudden switch of the pulsed laser, demonstrating the required responce time towards a new steady state. While turning to a few atoms with strong interactions both the optimal dressing probability and the response time are improved due to the collective feature in the blockade regime.

Remarkably our scheme provides a fresh way to save the laser powers (only depend on the relative strength) in dressing ground atoms to Rydberg levels, facilitating a robust and efficient Rydberg dressed-atom generation with reduced laser powers and optimal response times. An extension to a few-atom ensemble promotes extra advantages in developing a more efficient excitation probability with deeply reduced rise time, widely broadening its applications in controllable many-body physics as well as the quantum control of multi-level drivings.

\acknowledgements

This work was supported by the NSFC under Grants No. 11474094, No. 11104076, the ``Fundamental Research Funds for the Central Universities'', the Academic Competence Funds for the outstanding doctoral students under YBNLTS2019-023, the
Specialized Research Fund for the Doctoral Program of Higher Education No.
20110076120004.

\appendix

\section{Analytical steady-state solutions}

Here, we present a detailed derivation to calculate the analytical expressions of one-atom steady probe absorption and Rydberg dressing probability, in order to quantitatively compare with our numerical calculations. Given the Hamiltonian (\ref{Ham2}) and the master equation (\ref{maseq}) the optical Bloch equations for elements of one-atom density matrix can be written as
\begin{flalign}
&\dot{\rho}_{gg}=\Gamma\rho_{mm}-\Omega_{p}\rho_{gm}^I  \nonumber \\
&\dot{\rho}_{mm}=-\Gamma(\rho_{mm}+\rho_{ee})-\Omega_{1}\rho_{me}^I+\Omega_{p}\rho_{gm}^I  \nonumber\\
&\dot{\rho}_{ee}=-\Gamma\rho_{ee}+\gamma\rho_{rr}-\Omega_{20}\rho_{er}^I+\Omega_{1}\rho_{me}^I   \nonumber\\
&\dot{\rho}_{rr}=-\gamma\rho_{rr}+\Omega_{20}\rho_{er}^I  \nonumber\\
&\dot{\rho}_{gm}=(-\dfrac{\Gamma}{2}-i\Delta_{m})\rho_{gm}+\dfrac{i\Omega_{p}}{2}(\rho_{gg}-\rho_{mm})+\frac{i\Omega_{1}}{2}\rho_{ge}  \nonumber\\
&\dot{\rho}_{ge}=(-\dfrac{\Gamma}{2}-i\Delta_{m})\rho_{ge}+\dfrac{i\Omega_{1}}{2}\rho_{gm}+\frac{i\Omega_{20}}{2}\rho_{gr}-\frac{i\Omega_{p}}{2}\rho_{me}   \nonumber \\
&\dot{\rho}_{gr}=(-\dfrac{\gamma}{2}-i\Delta_{m})\rho_{gr}+\dfrac{i\Omega_{20}}{2}\rho_{ge}-\frac{i\Omega_{p}}{2}\rho_{mr}  \nonumber\\
&\dot{\rho}_{me}=-\Gamma\rho_{me}+\dfrac{i\Omega_{1}}{2}(\rho_{mm}-\rho_{ee})+\dfrac{i\Omega_{20}}{2}\rho_{mr}-\frac{i\Omega_{p}}{2}\rho_{ge}  \nonumber\\
&\dot{\rho}_{mr}=(-\dfrac{\Gamma+\gamma}{2})\rho_{mr}-\dfrac{i\Omega_{1}}{2}\rho_{er}+\frac{i\Omega_{20}}{2}\rho_{me}-\frac{i\Omega_{p}}{2}\rho_{gr}  \nonumber\\
&\dot{\rho}_{er}=(-\dfrac{\Gamma+\gamma}{2})\rho_{er}-\dfrac{i\Omega_{1}}{2}\rho_{mr}+\frac{i\Omega_{20}}{2}(\rho_{ee}-\rho_{rr}) \end{flalign}
where the conserved population is $\rho_{gg}+\rho_{mm}+\rho_{ee}+\rho_{rr}=1$. With the steady-state assumption $\dot{\rho}_{ij}=0$ all steady state solutions $\rho_{ij}^s$ are analytically solvable, part of which are presented here, scaled by the decay rate $\Gamma$.

The probe absorption and dispersion can be respectively obtained from the imaginary and real part of $\rho_{gm}^s$ as shown in Eq.(6), described by
\begin{widetext}
\begin{flalign}
&\frac{(\rho_{gm}^{I})^s}{\Omega_p}=\frac{\gamma^{2}(1+4\Delta_{m}^{2}+\Omega_{1}^{2})+\gamma(2+\Omega_{1}^{2})\Omega_{20}^{2}+[4\Delta_{m}^{2}+16\Delta_{m}^{4}+\Omega_{20}^{4}+4\Delta_{m}^{2}(\Omega_{1}^{2}-2\Omega_{20}^{2})]}{4\Delta_{m}^{2}+\gamma^{2}(1+\Omega_{1}^{2})^{2}+2\gamma(1+\Omega_{1}^{2})\Omega_{20}^{2}+4\Delta_{m}^{2}(\Omega_{1}^{2}\Omega_{20}^{2}-4\Delta_{m}^{2})^{2}[32\Delta_{m}^{4}+\Omega_{20}^{4}+8\Delta_{m}^{2}(\Omega_{1}^{2}-\Omega_{20}^{2})]}  \nonumber\\
&\frac{(\rho_{gm}^{R})^s}{\Omega_p}=\frac{2\Delta_{m}[4\Delta_{m}^{2}+\gamma^{2}(1-\Omega_{1}^{2})+2\gamma\Omega_{20}^{2}+(4\Delta_{m}^{2}-\Omega_{20}^{2})(4\Delta_{m}^{2}-\Omega_{1}^{2}-\Omega_{20}^{2})]}{4\Delta_{m}^{2}+\gamma^{2}(1+\Omega_{1}^{2})^{2}+2\gamma(1+\Omega_{1}^{2})\Omega_{20}^{2}+4\Delta_{m}^{2}(\Omega_{1}^{2}\Omega_{20}^{2}-4\Delta_{m}^{2})^{2}[32\Delta_{m}^{4}+\Omega_{20}^{4}+8\Delta_{m}^{2}(\Omega_{1}^{2}-\Omega_{20}^{2})]}
\end{flalign}
\end{widetext}

Noting that $(\rho_{gm}^{I(R)})^s \propto \Omega_p$ arising that the strengths of probe absorption and dispersion linearly increase with the probe laser intensity. By substituting $\rho_{gm}^s$ [Eq.(\ref{gm})] into (A1) and setting $\dot{\rho}_{ij}=0$ we are able to solve a complex expression for the steady dressing probability of Rydberg state under the first-order perturbations, labeled by $P_{1,r}^{on}$ in the maintext, which is given by
\begin{widetext}
\begin{eqnarray}
\rho_{rr}^s&=&\frac{\Omega_{1}^{2}\Omega_{20}^{2}(2+\Omega_{1}^{2}+\Omega_{20}^{2})\Omega_{p}^{2}}{\gamma^{2}(\Omega_{1}^{2}+1)^{3}+\Omega_{20}^{4}(2+\Omega_{1}^{2}+\Omega_{20}^{2})+\Omega_{20}^{2}[2+5\Omega_{1}^{2}+2(\Omega_{1}^{4}+\Omega_{20}^{4})]\Omega_{p}^{2}+\gamma\{2\Omega_{20}^{2}[2+\Omega_{1}^{4}+(3\Omega_{1}^{2}+2\Omega_{20}^{2})]+(2+\Omega_{1}^{2})^{2}\Omega_{p}^{2}\}} \\
\rho_{rr}^s&=&\frac{\Omega_{1}^{2}\Omega_{20}^{2}\Omega_{p}^{2}[2\gamma+3\gamma\Omega_{1}^{2}+2\Omega_{1}^{4}+(2+3\Omega_{1}^{2})\Omega_{20}^{2}+\Omega_{20}^{4}]}{[\gamma(1+\Omega_{1}^{2})(2+\Omega_{1}^{2})+\Omega_{20}^{2}(2+\Omega_{1}^{2}+\Omega_{20}^{2})][(\Omega_{1}^{2}+\Omega_{20}^{2})+(2\Omega_{1}^{2}+\Omega_{20}^{2})^{2}]}     
\end{eqnarray}
\end{widetext}
 with respect to $\Delta_m=0$ and $\Delta_m=\pm\Omega_{eff}/2$, accordingly. Due to the complexity these analytical expressions are plotted in Fig.\ref{eigsp}(b-d) by circles, as compared to our numerical results. A perfect agreement is revealed strongly supporting the correctness of our calculations. Besides by doing derivation of Eq.(A3) to the ratio $\Omega_{20}/\Omega_1$ it arrives to an approximated equation $\Omega_{20}/\Omega_1\approx \sqrt{\gamma/\Gamma}$ by which a maximal steady probability $P_{1,r}^{opt}$ can be attainable.

\section{Comparing response speeds of probe absorption and Rydberg dressing}

\begin{figure}
\includegraphics[width=3.3in,height=3.4in]{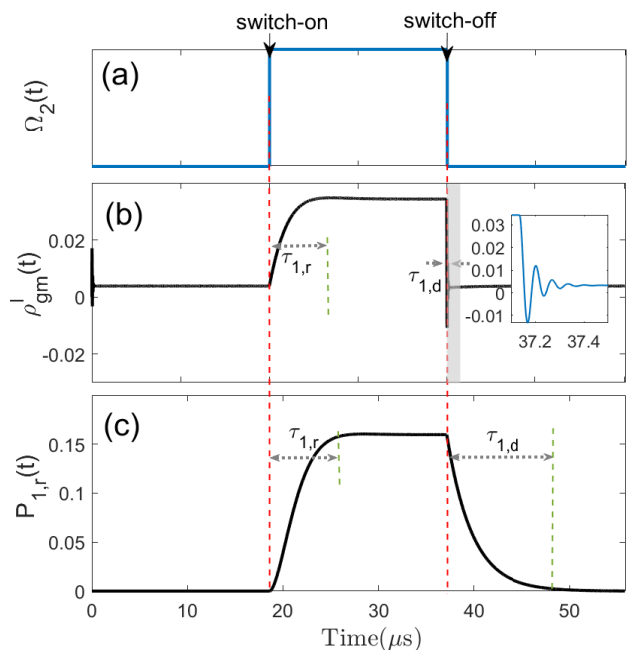}
\caption{A comparison for transient response speeds including the rise time $\tau_{1,r}$ and the fall time $\tau_{1,d}$ of the probe absorption $\rho_{gm}^I(t)$ and the Rydberg dressing population $P_{1,r}(t)$ in the one-atom frame. (a) The shape of time-dependent addressing pulse $\Omega_2(t)$. (b) and (c) show practical dynamics of the probe absorption $\rho_{gm}^I(t)$ and the Rydberg dressing probability $P_{1,r}(t)$. Inset of (b) is an enlarged picture describing a fast damped oscillation to the probe absorption response by a sudden switch-off of the addressing laser. Parameters are same as used in Fig.\ref{realp}(a2).}
\label{comresponse}
\end{figure}

\begin{table}[tbp]
\begin{tabular}{c|cc}
\hline\hline
measured variables &  $ \tau_{1,r}(\mu s)$ & $\tau_{1,d}(\mu s)$ \\
 \hline 
$\rho_{gm}^I(t)$ & 4.218 & 0.292  \\ 
$P_{1,r}(t)$ & 6.870 & 10.398 \\ 
\hline\hline
\end{tabular}%
\caption{ By switching on or off the addressing laser $\Omega_2(t)$ a quantitative comparison for the transient response speeds of $\rho_{gm}^I(t)$ and $P_{1,r}(t)$ are given, corresponding to the results in Fig.\ref{comresponse}.}%
\label{smm}
\end{table}

In this appendix, we numerically study different transient responses for the probe absorption $\rho_{gm}^I(t)$ as well as the Rydberg dressing probability $P_{1,r}(t)$ by doing a sudden switch of $\Omega_2(t)$, demonstrating that the required time for the atomic dressing is longer than for the probe absorption. The numerical search algorithm is carried out by tracking the minimal time required towards to a new steady state through suddenly switch-on or switch-off of the control pulse.

Figure \ref{comresponse}(a) shows the shape of an one-period control pulse $\Omega_2(t)$ where the operations of switch-on or switch-off occur at $t=18.57\mu$s and $37.14\mu$s, respectively. With the incidence of such a pulse we numerically calculate the realistic dynamics of $\rho_{gm}^I(t)$ and $P_{1,r}(t)$ as displayed in Fig. \ref{comresponse}(b-c), where the rise and fall times are importantly denoted. 
In general it is observed that both $\rho_{gm}^I(t)$ and $P_{1,r}(t)$ represent in-phase evolutions towards its new steady states with respect to $\Omega_2(t)$, however accompanied by quite different response times. A quantitative comparison of response times is given in Table \ref{smm} presenting that the rise and fall times of $P_{1,r}(t)$ is relatively longer, caused by a small decay rate of highly-excited Rydberg state.  

In addition it is notable that the fall time $\tau_{1,d}$ of $\rho_{gm}^I(t)$ is significantly reduced by more than one order of magnitude, accordingly in the inset of Fig. \ref{comresponse}(b) the behavior $\rho_{gm}^I(t)$ has an ultra-fast damped oscillation to be stable. The reason for that is due to the relaxation process in a driven-dissipative system, quantitatively agreeing with the timescale as found in Ref.\cite{Zhang18}. Therefore we conclude that a practical operation on preparing Rydberg-dressed atoms requires a timescale of $(1\sim 10)\mu$s, longer than the time for recording the transient response of probe laser transmission in EIT environment which typically needs $(0.1\sim1.0)\mu$s by experimentalists \cite{Dmochowski16}.


\bigskip

\begin{thebibliography}{99}

\bibitem{Johnson10} J. Johnson and S. Rolston, ``Interactions between Rydberg-dressed atoms'', Phys. Rev. A \textbf{82}, 033412 (2010).

\bibitem{Saffman16} M. Saffman, ``Quantum computing with atomic qubits and Rydberg interactions: progress and challenges'', J. Phys. B, \textbf{49}, 202001 (2016). 

\bibitem{Honer10} J. Honer, H. Weimer, T. Pfau, and H. B\"{u}chler, ``Collective Many-Body Interaction in Rydberg Dressed Atoms'', Phys. Rev. Letts. \textbf{105}, 160404 (2010).

\bibitem{Pupillo10} G. Pupillo, A. Micheli, M. Boninsegni, I. Lesanovsky, and P. Zoller, ``Strongly Correlated Gases of Rydberg-Dressed Atoms: Quantum and Classical Dynamics'', Phys. Rev. Letts. \textbf{104}, 223002 (2010).

\bibitem{Henkel10} N. Henkel, R. Nath, and T. Pohl, ``Three-Dimensional Roton Excitations and Supersolid Formation in Rydberg-Excited Bose-Einstein Condensates'', Phys. Rev. Letts. \textbf{104}, 195302 (2010). 

\bibitem{Balewski14} J. Balewski, A. Krupp, A. Gaj, S. Hofferberth, .R. L\"{o}w and T. Pfau, ``Rydberg dressing: understanding of collective many-body effects and implications for experiments'', New J. Phys. \textbf{16}, 063012 (2014).

\bibitem{Bijnen15} R. Bijnen and T. Pohl, ``Quantum Magnetism and Topological Ordering via Rydberg Dressing near F\"{o}rster Resonances'', Phys. Rev. Letts. \textbf{114}, 243002 (2015).

\bibitem{Glaetzle15} A. Glaetzle, M. Dalmonte, R. Nath, C. Gross, I. Bloch, and P. Zoller, ``Designing Frustrated Quantum Magnets with Laser-Dressed Rydberg Atoms'', Phys. Rev. Letts. \textbf{114}, 173002 (2015).

\bibitem{Keating13} T. Keating, K. Goyal, Y. Jau, G. Biedermann, A. Landahl, and I. Deutsch, ``Adiabatic quantum computation with Rydberg-dressed atoms'', Phys. Rev. A \textbf{87}, 052314 (2013).

\bibitem{Keating15} T. Keating, R. Cook, A. Hankin, Y. Jau, G. Biedermann, and I. Deutsch, ``Robust quantum logic in neutral atoms via adiabatic Rydberg dressing'', Phys. Rev. A \textbf{91}, 012337 (2015).

\bibitem{Jau16} Y. Jau, A. Hankin, T. Keating, I. Deutsch, G. Biedermann, ``Entangling atomic spins with a Rydberg-dressed spin-flip blockade'', Nat. Phys. \textbf{12}, 71 (2016).

\bibitem{Mohapatra07} A. Mohapatra, T. Jackson, and C. Adams, ``Coherent Optical Detection of Highly Excited Rydberg States Using Electromagnetically Induced Transparency'', Phys. Rev. Letts. \textbf{98}, 113003 (2007).

\bibitem{Pritchard10} J. Pritchard, D. Maxwell, A. Gauguet, K. Weatherill, M. Jones, and C. Adams, ``Cooperative Atom-Light Interaction in a Blockaded Rydberg Ensemble'', Phys. Rev. Letts. \textbf{105}, 193603 (2010).

\bibitem{Petrosyan11} D. Petrosyan, J. Otterbach, and M. Fleischhauer, ``Electromagnetically Induced Transparency with Rydberg Atoms'', Phys. Rev. Letts. \textbf{107}, 213601 (2011).

\bibitem{Gaul16} C. Gaul, B. DeSalvo, J. Aman, F. Dunning, T. Killian, and T. Pohl, ``Resonant Rydberg Dressing of Alkaline-Earth Atoms via Electromagnetically Induced Transparency'', Phys. Rev. Letts. \textbf{116}, 243001 (2016).

\bibitem{Firstenberg16} O. Firstenberg, C. Adams and S. Hofferberth,``Nonlinear quantum optics mediated by Rydberg interactions'', J. Phys. B \textbf{49}, 152003 (2016).

\bibitem{Vogt18} T. Vogt, C. Gross, T. Gallagher, and W. Li, ``Microwave-assisted Rydberg electromagnetically induced transparency'', Opt. Lett. \textbf{43}, 1822 (2018).

\bibitem{Leseleuc17} S. L\'{e}s\'{e}leuc, D. Barredo, V. Lienhard, A. Browaeys, and T. Lahaye, ``Optical Control of the Resonant Dipole-Dipole Interaction between Rydberg Atoms'', Phys. Rev. Letts. \textbf{119}, 053202 (2017).

\bibitem{Carr12} C. Carr, M. Tanasittikosol, A. Sargsyan, D. Sarkisyan, C. Adams and K. Weatherill, ``Three-photon electromagnetically induced transparency using Rydberg states'', Opt. Lett. \textbf{37}, 3858 (2012).

\bibitem{Agarwal96} G. Agarwal and W. Harshawardhan, ``Inhibition and Enhancement of Two Photon Absorption'', Phys. Rev. Letts. \textbf{77}, 1039 (1996).

\bibitem{Harris98} S. Harris and Y. Yamamoto, ``Photon Switching by Quantum Interference'', Phys. Rev. Letts. \textbf{81}, 3611 (1998).

\bibitem{Mulchan00} N. Mulchan, D. Ducreay, R. Pina, M. Yan and Y. Zhu, ``Nonlinear excitation by quantum interference in a Doppler-broadened rubidium atomic system'', J. Opt. Am. B, \textbf{17}, 820 (2000).

\bibitem{Dutta07} B. Dutta and P. Mahapatra, ``Nonlinear optical effects in a doubly driven four-level atom'', Phys. Scr. \textbf{75}, 345 (2007).

\bibitem{Che14} J. Che, H. Zheng, Z. Zhang, X. Yao, C. Li, Z. Wu and Y. Zhang, ``Rydberg dressing evolution via Rabi frequency control in thermal atomic vapors'', Phys. Chem. Chem. Phys. \textbf{16}, 18840 (2014).

\bibitem{Kondo15} J. Kondo, N. Sibalic, A. Guttridge, C. Wade, N. Melo, C. Adams, and K. Weatherill, ``Observation of interference effects via four-photon excitation of highly excited Rydberg states in thermal cesium vapor'', Opt. Lett. \textbf{40}, 5570 (2015).

\bibitem{Veit16} C. Veit, G. Epple, H. K\"{u}bler, T. Euser, P. Russell and R. L\"{o}w, ``RF-dressed Rydberg atoms in hollow-core fibres'', J. Phys. B, \textbf{49}, 134005 (2016).


\bibitem{Zhang18} Q. Zhang, Z. Bai and G. Huang, ``Fast-responding property of electromagnetically induced transparency in Rydberg atoms'', Phys. Rev. A \textbf{97}, 043821 (2018).

\bibitem{Ate11} C. Ates, S. Sevincli, and T. Pohl, ``Electromagnetically induced transparency in strongly interacting Rydberg gases'', Phys. Rev. A \textbf{83}, 041802R (2011).
 
\bibitem{Xu16} W. Xu and B. DeMarco, ``Velocity-selective electromagnetically-induced-transparency measurements of potassium Rydberg states'', Phys. Rev. A \textbf{93}, 011801R (2016).
 

\bibitem{Sheng17} J. Sheng, Y. Chao, S. Kumar, H. Fan, J. Sedlacek, and J. Shaffer, ``Intracavity Rydberg-atom electromagnetically induced transparency using a high-finesse optical cavity'', Phys. Rev. A \textbf{96}, 033813 (2017).

\bibitem{Jiao17} Y. Jiao, L. Hao, X. Han, S. Bai, G. Raithel, J. Zhao, and S. Jia, ``Atom-Based Radio-Frequency Field Calibration and Polarization Measurement Using Cesium $nD_J$ Floquet States'', Phys. Rev. Applied \textbf{8}, 014028 (2017).

\bibitem{Rawat18} H. Rawat, S. Dubey and V. Ojha, ``Distinction between double electromagnetically induced transparency and double Autler-Townes splitting in RF-driven four-level ladder $^{87}$Rb atomic vapor'', J. Phys. B \textbf{51}, 155401 (2018).

\bibitem{Chai17} X. Chai, L. Zhang, D. Ma, L. Yan, H. Bao, and J. Qian, ``Anomalous excitation enhancement with Rydberg-dressed atoms'', Phys. Rev. A \textbf{96}, 053417 (2017).

\bibitem{Baur14} S. Baur, D. Tiarks, G. Rempe, and S. D\"{u}rr, ``Single-Photon Switch Based on Rydberg Blockade'', Phys. Rev. Letts. \textbf{112}, 073901 (2014).

\bibitem{Gorniaczyk14} H. Gorniaczyk, C. Tresp, J. Schmidt, H. Fedder, and S. Hofferberth, ``Single-Photon Transistor Mediated by Interstate Rydberg Interactions'', Phys. Rev. Letts. \textbf{113}, 053601 (2014).

\bibitem{Li15} W. Li and I. Lesanovsky, ``Coherence in a cold-atom photon switch'', Phys. Rev. A \textbf{92}, 043828 (2015).

\bibitem{Qian17} J. Qian, ``Robust quantum switch with Rydberg excitations'', Sci. Reps. \textbf{7}, 12952 (2017).

\bibitem{Zhang09} L. Zhang, Z. Feng, A. Li, J. Zhao, C. Li and S. Jia, ``Measurement of quantum defects of nS and nD states using field ionization spectroscopy in ultracold cesium atoms'', Chin. Phys. B \textbf{18}, 1838 (2009).


\bibitem{Urban09} E. Urban, T. Johnson, T. Henage, L. Isenhower, D. Yavuz, T. Walker and M. Saffman, ``Observation of Rydberg blockade between two atoms'', Nat. Phys. \textbf{5}, 110(2009).

\bibitem{Gaetan09} A. Ga\"{e}tan, Y. Miroshnychenko, T. Wilk, A. Chotia, M. Viteau, D. Comparat, P. Pillet, A. Browaeys and P. Grangier, ``Observation of collective excitation of two individual atoms in the Rydberg blockade regime'', Nat. Phys. \textbf{5}, 115(2009).

\bibitem{Wilk10} T. Wilk, A. Ga\"{e}tan, C. Evellin, J. Wolters, Y. Miroshnychenko, P. Grangier, and A. Browaeys, ``Entanglement of Two Individual Neutral Atoms Using Rydberg Blockade'', Phys. Rev. Letts. \textbf{104}, 010502 (2010).


\bibitem{Heidemann07} R. Heidemann, U. Raitzsch, V. Bendkowsky, B. Butscher, R. L\"{o}w, L. Santos, and T. Pfau, ``Evidence for Coherent Collective Rydberg Excitation in the Strong Blockade Regime'', Phys. Rev. Letts. \textbf{99}, 163601(2007).


\bibitem{Beguin13} L. B\'{e}guin, A. Vernier, R. Chicireanu, T. Lahaye, and A. Browaeys, ``Direct Measurement of the van der Waals Interaction between Two Rydberg Atoms'', Phys. Rev. Letts. \textbf{110}, 263201(2013).

\bibitem{Qian16} J. Qian, ``Resonance-enhanced collective effect in a triangle arrangement of Rydberg atoms with anisotropic interactions'', J. Opt. Soc. Am. B, \textbf{33}, 1749(2016).

\bibitem{Ma19} D. Ma, K. Zhang and J. Qian, ``Properties of collective Rabi oscillations with two Rydberg atoms'', Chin. Phys. B \textbf{28}, 013204(2019).

\bibitem{Beterov09} I. Beterov, I. Ryabtsev, D. Tretyakov, and V. Entin, ``Quasiclassical calculations of blackbody-radiation-induced depopulation rates and effective lifetimes of Rydberg $nS$, $nP$, and $nD$ alkali-metal atoms with $n\leq80$'', Phys. Rev. A \textbf{79}, 052504 (2009).

\bibitem{Zoubi15} H. Zoubi, ``Van der Waals interactions among alkali Rydberg atoms with excitonic states'', J. Phys. B, \textbf{48}, 185002(2015).

\bibitem{Johnson08} T. Johnson, E. Urban, T. Henage, L. Isenhower, D. Yavuz, T. Walker, and M. Saffman, ``Rabi Oscillations between Ground and Rydberg States with Dipole-Dipole Atomic Interactions'', Phys. Rev. Letts. \textbf{100}, 113003(2008). 

\bibitem{Dudin12} Y. Dudin and A. Kuzmich, ``Strongly Interacting Rydberg Excitations of a Cold Atomic Gas'', Science, \textbf{336}, 887(2012). 

\bibitem{Zeiher15} J. Zeiher, P. Schauß, S. Hild, T. Macrì, I. Bloch, and C. Gross, ``Microscopic Characterization of Scalable Coherent Rydberg Superatoms'', Phys. Rev. X \textbf{5}, 031015(2015). 

\bibitem{Sibalic16} N. Sibalic, J. Kondo, C. Adams and K. Weatherill, ``Dressed-state electromagnetically induced transparency for light storage in uniform-phase spin waves'', Phys. Rev. A \textbf{94}, 033840(2016). 


\bibitem{Dmochowski16} G. Dmochowski, A. Feizpour, M. Hallaji, C. Zhuang, A. Hayat, and A. Steinberg, ``Experimental Demonstration of the Effectiveness of Electromagnetically Induced Transparency for Enhancing Cross-Phase Modulation in the Short-Pulse Regime'', Phys. Rev. Letts. \textbf{116}, 173002(2016). 





















































































































































































































 





















%
%
%
%
%
%
%
%
%
%
%
%
%
%
%
%
%
%
%
%
%
%
%
%
%
%
%
%
%
%
%




























\end{thebibliography}
\end{document}